\documentstyle[11pt,epsfig]{article}
\date{}
\begin{document}
\title{{\bf Noether symmetric minisuperspace model of $f(R)$ cosmology}}
\author{Babak Vakili\thanks{%
email: b-vakili@iauc.ac.ir}\\\\
{\small {\it Department of Physics, Azad University of Chalous,}}\\{\small {\it P. O. Box 46615-397, Chalous, Iran}}}
\maketitle
\begin{abstract} We study the metric $f(R)$ cosmology using Noether symmetry approach by utilizing the behavior of the corresponding Lagrangian under infinitesimal generators of the desired symmetry. The existence of Noether symmetry of the cosmological $f(R)$ minisuperspace helps us to find out the form of $f(R)$ function for which such symmetry exist. It is shown that the resulting form for $f(R)$ yields a power law expansion for the cosmic scale factor. We also show that in the corresponding Noether symmetric quantum model, the solutions to the Wheeler-DeWitt equation can be expressed as a superposition of states of the form $e^{iS}$. It is shown that in terms of such wavefunctions the classical trajectories can be recovered.\vspace{5mm}\newline PACS numbers: 04.20.Fy, 04.50.+h, 98.80.Qc
\end{abstract}

\section{Introduction}
In this letter we consider the vacuum universe in the flat FRW space-time in the framework of the metric formalism of $f(R)$ gravity \cite{1}, and by following the so-called Noether symmetry approach \cite{4}, \cite{S}. We look for the form of the function $f(R)$ which the Lagrangian admits the desired Noether symmetry. To do this, we shall make up a point-like Lagrangian in which the scale factor $a$ and the Ricci scalar $R$ play the role of independent dynamical variables. We shall see that by demanding the Noether symmetry as a feature of the Lagrangian, we can obtain the explicit form of the function $f(R)$ which as we will show yields a power law expansion for the corresponding classical cosmology. Also, we shall deal with the quantization of the model and show that applying the quantum version of the Noether symmetry on the wavefunction of the universe (which satisfies the Wheeler- DeWitt equation) results an oscillatory behavior for the wavefunction in the direction of the symmetry. This property together with the semiclassical approximation for the canonical quantum cosmology help us to recover the classical trajectory and to show that the cosmological scale factor obeys a power law expansion.
\section{The Noether symmetric $f(R)$ model}
We start from the $(n+1)$-dimensional action
\begin{equation}\label{A}
{\cal S}=\int d^{n+1}x\sqrt{-g}f(R),
\end{equation}where $R$ is the scalar curvature and $f(R)$ is an arbitrary
function of $R$. Since our goal is to study models
which exhibit Noether symmetry, we do not include any matter
contribution in the action. We
assume that the geometry of space-time is described by the flat
FRW metric
\begin{equation}\label{B}
ds^2=-dt^2+a^2(t)\sum_{i=1}^{n}(dx^i)^2.
\end{equation}
In order to apply the Noether symmetry approach, we have to write the action in the point-like form ${\cal S}=\int dt {\cal L}(a,\dot{a},R,\dot{R})$,
in which the scale factor $a$ and scalar curvature $R$ play the role of independent dynamical variables. It can be easily verified that for the FRW metric (\ref{B}), such point-like Lagrangian takes the form
\begin{equation}\label{C}
{\cal L}(a,\dot{a},R,
\dot{R})=n(n-1)\dot{a}^2a^{n-2}f'+2n\dot{a}\dot{R}a^{n-1}f''+a^n(f'R-f),
\end{equation}where dot and prime represent derivative with respect to $t$ and $R$ respectively. Also, we have the zero energy condition (Hamiltonian constraint) associated with the above Lagrangian as
\begin{equation}\label{D}
E_{\cal L}=n(n-1)\dot{a}^2a^{n-2}f'+2n\dot{a}\dot{R}a^{n-1}f''-a^n(f'R-f)=0.
\end{equation} Now, let us introduce the Noether symmetry induced on the model by a vector field
$X$ on the tangent space $TQ=\left({\bf q},\dot{{\bf q}}\right)$ of
the configuration space ${\bf q}=\left(a,R\right)$ of Lagrangian
(\ref{C}) through
\begin{equation}\label{E}
X=A^i({\bf q}) \frac{\partial}{\partial q^i}+\frac{d
A^i({\bf q})}{dt}\frac{\partial}{\partial \dot{q}^i},
\end{equation}where $A^i(i=1,2)=(\alpha, \beta)$ are unknown functions of $a$ and $R$. The Noether symmetry implies that the Lie derivative of the Lagrangian with respect to this vector field vanishes, that is, $L_X {\cal L}=0$, which leads the following system of partial differential equation
\begin{eqnarray}\label{F}
\left\{
\begin{array}{ll}
n(f'R-f)\alpha +a f''R\beta =0,\\
f''\frac{\partial \alpha}{\partial R}=0,\\
(n-1)(n-2)f'\alpha + (n-1)a f''\beta +2(n-1)af'\frac{\partial
\alpha}{\partial a}+2a^2f''\frac{\partial \beta}{\partial
a}=0,\\
(n-1)f''\alpha+af'''\beta+af''\frac{\partial
\alpha}{\partial a}+(n-1)f'\frac{\partial \alpha}{\partial R}+a
f''\frac{\partial \beta}{\partial R}=0.
\end{array}
\right.
\end{eqnarray} It easy to see that \cite{5}, in the case where $f''=0$, we are led to the Minkowski solution for the equations of motion, which is not of interest in a cosmological point of view. Therefore, we remove the case $f''=0$ from our consideration. When $f'' \neq 0$ the system (\ref{F}) admits the following solutions \cite{5}
\begin{equation}\label{G}
\alpha(a)=-\frac{2}{n+1}a^{-\frac{n-1}{2}},\hspace{.5cm}\beta(a,R)=Ra^{-\frac{n+1}{2}},
\end{equation}
and
\begin{equation}\label{H}
f(R)=R^{\frac{2n}{n+1}}.
\end{equation}This means that in the context of the metric formalism of $f(R)$ cosmology, a flat FRW metric
has Noether symmetry if the corresponding action is given by equation (\ref{H}).
\section{Classical cosmology}
Noether symmetry approach is a powerful tool in
finding the solution to a given Lagrangian, including the one
presented above. In this approach, one is concerned with finding
the cyclic variables related to conserved quantities and
consequently reducing the dynamics of the system to a manageable
one. The existence of Noether symmetry means that there exists a constant of motion. Indeed, from (\ref{E}) we have
\begin{equation}\label{I}
L_X{\cal L}=A^i({\bf q})\frac{\partial{\cal L}}{\partial q^i}+\frac{d A^i({\bf q})}{dt}\frac{\partial {\cal L}}{\partial \dot{q}^i}=0.\end{equation}
Noting that $p_q=\frac{\partial {\cal L}}{\partial \dot{q}}$ (the momentum conjugate to $q$) and $\frac{\partial {\cal
L}}{\partial q}=\frac{dp_q}{dt}$, we get $\frac{d}{dt}A^i({\bf q})p_{q_i}=0$, which means that the constants of motion are
\begin{equation}\label{J}
Q=A^i({\bf q})p_{q_i}.\end{equation} To obtain the corresponding cosmology resulting from  $f(R)$ model (\ref{H}), we note that the momenta conjugate to variables $a$ and $R$ are
\begin{equation}\label{K}
p_a=\frac{\partial {\cal L}}{\partial
\dot{a}}=2n(n-1)\dot{a}a^{n-2}f'+2na^{n-1}\dot{R}f'',\hspace{0.5cm}p_R=\frac{\partial {\cal L}}{\partial
\dot{R}}=2n\dot{a}a^{n-1}f''.\end{equation} Using these equations and the zero energy condition (\ref{D}), it is easy to show that
\begin{equation}\label{L}
Q=-\frac{8n^2}{(n+1)^2}\frac{d}{dt}\left(a^{\frac{n-1}{2}}R^{\frac{n-1}{n+1}}\right),\hspace{0.5cm}
2n^2\dot{a}^2R^{-1}+\frac{4n^2}{n+1}\dot{a}\dot{R}aR^{-2}=a^2,
\end{equation}which can be immediately integrated with the result (assuming
$a(t=0)=0$)
\begin{equation}\label{M}
a(t) \sim t^{\frac{2(3n-1)}{n^2-1}}.
\end{equation}Therefore, the solution for the cosmic scale factor which is obtained from Noether symmetry represents a power law expansion. It
is remarkable from (\ref{M}) that models with spatial dimension $n\leq 5$
obey an accelerated power law expansion while for $n>5$ a
decelerated expansion occurs.

Now, we want to see that whether the above model can provide gravitational alternative for dark energy or not. For this purpose one can write the effective energy density and pressure for our $f(R)$ model as follows \cite{5}
\begin{equation}\label{N}
\rho=\frac{1}{2}\left(Rf'-f\right)-nH\dot{R}f'',\hspace{0.5cm}
P=f'''\dot{R}^2+(n-1)H\dot{R}f''+\ddot{R}f''+\frac{1}{2}\left(f-Rf'\right),
\end{equation}where $H=\dot{a}/a$. Thus, we can define the effective equation of state (EoS) parameter $\omega_{eff}$ as
\begin{equation}\label{O}
w_{eff}=\frac{P}{\rho}=\frac{f'''\dot{R}^2+(n-1)H\dot{R}f''+\ddot{R}f''+\frac{1}{2}\left(f-Rf'\right)}
{\frac{1}{2}\left(Rf'-f\right)-nH\dot{R}f''}.
\end{equation}Therefore, substituting (\ref{H}) and (\ref{M}), after some algebra we get the following EoS parameter
\begin{equation}\label{P}
w_{eff}=-1+\frac{(n-1)(n+1)}{n(3n-1)}.
\end{equation}Since $-1<w_{eff}<0$, one type of dark energy, the so-called quintessence, can be addressed in the model described above while another type of dark energy known as phantom, with $\omega_{eff}<-1$, cannot be accounted for.
\section{Quantum cosmology}
The study of quantum cosmology of the model presented above is the goal we shall pursue
in this section. For this purpose we construct the Hamiltonian of our model. From now on we consider the case $n=3$. Substituting (\ref{H}) into (\ref{C}), we obtain the Lagrangian and Hamiltonian of the Noether symmetric model as
\begin{equation}\label{Q}
{\cal L}=9\dot{a}^2aR^{1/2}+\frac{9}{2}\dot{a}\dot{R}a^2R^{-1/2}+\frac{1}{2}a^3R^{3/2},\hspace{0.3cm}
{\cal H}=\frac{2}{9}a^{-2}R^{1/2}p_ap_R-\frac{4}{9}a^{-3}R^{3/2}p_R^2-\frac{1}{2}a^3R^{3/2}.\end{equation}
To transform the above Lagrangian to a more manageable
form, consider the following change of variables \cite{6}
\begin{equation}\label{R}
u(a,R)=-a^2+(aR^{1/2})^{\mu},\hspace{.5cm}v(a,R)=(aR^{1/2})^{\nu},\end{equation}where $\mu$ and $\nu$ are some constants. In terms of these new variables, the Hamiltonian takes the form
\begin{equation}\label{S}
{\cal H}=-\frac{10}{9}\mu v^{\frac{\mu-1}{\nu}}p_u^2-\frac{2}{3}\nu v^{\frac{\nu-1}{\nu}}p_u p_v -\frac{1}{2}v^{3/\nu}.\end{equation}
It is clear from this Hamiltonian that $u$ is cyclic and the Noether symmetry is given by $p_u=Q=\mbox{cons.}$. The quantum state of the universe is
then described by a wavefunction in the minisuperspace, satisfying the Wheeler-DeWitt
equation, that is, ${\cal H}\Psi = 0$, where ${\cal H}$ is the operator form of the
Hamiltonian given by equation (\ref{S}) and $\Psi$ is the wavefunction of the universe. On the other hand, the existence of a Noether symmetry
in the model reduces the dynamics through $p_u=Q$, where its quantum version can be considered as a constraint $p_u \Psi =Q \Psi$. Therefore, with the replacement $p_u\rightarrow -i\frac{\partial}{\partial u}$ and similarly for
$p_v$, the quantum cosmology of our Noether symmetric $f(R)$ model can be described by the following equations
\begin{equation}\label{T}
\left[\frac{10}{9}\mu v^{\frac{\mu-1}{\nu}}\frac{\partial^2}{\partial u^2}+\frac{\nu-1}{3}v^{-1/\nu}\frac{\partial}{\partial u}+
\frac{2}{3}\nu v^{\frac{\nu-1}{\nu}}\frac{\partial^2}{\partial u \partial v}-\frac{1}{2}v^{3/\nu}\right]\Psi(u,v)=0,\end{equation}
\begin{equation}\label{U}
-i\frac{\partial}{\partial u}\Psi(u,v)=Q\Psi(u,v).\end{equation}The eigenfunctions of the equations (\ref{T}) and (\ref{U}) can be written as
\begin{equation}\label{V}
\Psi_{\mu \nu}(u,v)=v^{\frac{1-\nu}{2\nu}}e^{i\left(Qu-\frac{5}{3}Qv^{\mu/\nu}-\frac{3}{16Q}v^{4/\nu}\right)}.\end{equation} The general solution to the quantum equations (\ref{T}) and (\ref{U}) may write as a superposition of the above eigenfunctions which are of the form $e^{iS}$. Therefore, by using of the semiclassical approximation for quantum cosmology, we can identify the exponential factor of (\ref{V}) with $S$, to recover the corresponding classical cosmology through $p_q=\frac{\partial S}{\partial q}$. This procedure results in
\begin{equation}\label{W}
v(t)=\left({\cal Q}t-t_0\right)^{\nu},\hspace{0.5cm}u(t)=\frac{1}{3}\left({\cal Q}t-t_0\right)^{\mu}-\frac{1}{108{\cal Q}^2}\left({\cal Q}t-t_0\right)^4,\end{equation}where where ${\cal Q}=-\frac{2}{9}Q$ and $t_0$ is an integrating constant. Going back to the variables $a$ and $R$, we obtain the corresponding classical cosmology as
\begin{equation}\label{X}
a(t)=\left[\frac{2}{3}\left({\cal Q}t-t_0\right)^{\mu}+\frac{1}{108{\cal Q}^2}\left({\cal Q}t-t_0\right)^4\right]^{1/2},\end{equation}
\begin{equation}\label{Y}
R(t)=\frac{\left({\cal Q}t-t_0\right)^2}{\frac{2}{3}\left({\cal Q}t-t_0\right)^{\mu}+\frac{1}{108{\cal Q}^2}\left({\cal Q}t-t_0\right)^4}.\end{equation}It is seen that in the late time, the universe evolves with a power law expansion and the scalar curvature goes to zero in this limit, which seems to be consistent with the present cosmological observations.
\section{Conclusions}
In this work, we have study $f(R)$ cosmology by Noether symmetry approach. This approach is based on the seek for Noether symmetry which allows one to find the form of the function $f(R)$. We have shown that the Noether symmetric model results in a power law expansion for the scale factor and can address the so-called quintessence type of dark energy. We have also quantized the model and shown that the wavefunction of the universe is of the form $e^{iS}$. By using the semiclassical approximation, we have recovered the classical trajectories.
\vspace{5mm}\newline \noindent {\bf
Acknowledgement}\vspace{2mm}\noindent\newline
I would like to express my special thanks to the
organizers of "Grasscosmofun'09" for their precious efforts of organization, and to the research council of Azad University of Chalous for
financial support.

\end{document}